\documentclass{cimento1}
\usepackage{amsmath}

\title{Leading isospin-breaking effects on the lattice \thanks{{\it Presented at ``XVI Incontri di Fisica delle Alte Energie'', Trieste (Italy), 19-21 April 2017.}}}
\author{D.~Giusti}
\instlist{
\inst
Dipartimento di Matematica e Fisica, Universit\`a degli Studi Roma Tre - Roma, Italy
\inst
INFN, Sezione di Roma Tre - Roma, Italy
}

\begin{document}

\maketitle

\begin{abstract}

Isospin is an almost exact symmetry of strong interactions and the corrections to the isosymmetric limit are, in general, at the percent level. For several hadronic quantities, such as pseudoscalar meson masses or the kaon leptonic and semileptonic decay rates, these effects are of the same order of magnitude of the errors quoted in nowadays lattice calculations and cannot be neglected any longer. In this talk I discuss some recent results for the pseudoscalar meson spectrum obtained by the RM123 Collaboration including isospin breaking corrections in first principles lattice simulations.

\end{abstract}

\section{Introduction}

The determination of several hadronic quantities relevant for flavour physics phenomenology using QCD simulations on the lattice has reached an impressive degree of precision such that both electromagnetic (e.m.) and strong isospin-breaking (IB) effects cannot be neglected anymore (see {\it e.g.} reference~\cite{ref:FLAG} and references therein).

In the past few years, using different methodologies, accurate lattice results including e.m. effects have been obtained for the hadron spectrum and for the leptonic decay rates of light pseudoscalar (PS) mesons~\cite{ref:Carrasco:2015xwa}. In reference~\cite{ref:deDivitiis:2013xla} the inclusion of QED effects in lattice QCD simulations has been carried out developing a method, the RM123 approach, which consists in an expansion of the lattice path-integral in powers of the two small parameters $ \alpha_{em}$ and $\left( m_d - m_u \right)$, where $\alpha_{em} \approx \left( m_d - m_u \right) / \Lambda_{QCD} \approx 1 \%$.

In this contribution I present the recent determinations of the pion, kaon and $D$-meson mass splittings computed by the RM123 Collaboration~\cite{ref:Giusti:2017dmp}.
The calculation has been performed within the quenched QED approximation ({\it i.e.} treating dynamical quarks as electrically neutral particles) and using the gauge ensembles generated by the European Twisted Mass Collaboration (ETMC) with $N_f = 2+1+1$ dynamical quarks.

\section{Evaluation of isospin-breaking corrections to pseudoscalar mesons}

The e.m.~and strong IB corrections to the mass of a PS meson can be written as 
 \begin{equation}
      M_{PS} = M^0_{PS} + \left[ \delta M_{PS} \right]^{QED} + \left[ \delta M_{PS} \right]^{QCD} 
      \label{e.MPS}
 \end{equation}
 with
  \begin{eqnarray}
        \left[ \delta M_{PS} \right]^{QED} & \equiv &  4 \pi \alpha_{em} \left[ \delta M_{PS} \right]^{em} + ... ~ , \label{e.MPS_QED}
        \\[2mm]
        \left[ \delta M_{PS} \right]^{QCD} & \equiv & \left( m_d - m_u \right) \left[ \delta M_{PS} \right]^{IB} + ... ~ , \label{e.MPS_QCD}
  \end{eqnarray}
where the ellipses stand for higher order powers of $\alpha_{em}$ and $\left( m_d - m_u \right)$, while $M^0_{PS}$ stands for the PS meson mass in the isosymmetric QCD theory.
The separation in equation~(\ref{e.MPS}) between the QED and QCD contributions, $\left[ \delta M_{PS} \right]^{QED}$ and $\left[ \delta M_{PS} \right]^{QCD}$, is renormalization scheme and scale dependent.

At first order in the expansion the pion mass splitting $\left (M_{\pi^+} - M_{\pi^0} \right)$ is a pure e.m.~effect.
All the disconnected diagrams generated by the sea quark charges cancel out in the difference $\left( M_{\pi^+} - M_{\pi^0} \right)$ and therefore this quantity is not affected by the quenched QED approximation. However, the determination of the pion mass splitting requires the evaluation of a disconnected diagram generated by valence quarks in the neutral pion which has been neglected, being numerically very noisy and computationally expensive. Since this diagram vanishes in the $SU(2)$ chiral limit~\cite{ref:deDivitiis:2013xla} and, consequently, it is of order of $O(\alpha_{em} m_{\ell})$, this contribution is expected to be numerically a small correction at the physical pion mass. At the physical pion mass and in the continuum and infinite volume limits for the pion mass splitting we obtained the result
\begin{eqnarray}
\label{e.pion_splitting}
    M_{\pi^+} - M_{\pi^0} & = & 4.21 ~ (23)_{stat+fit} ~ (13)_{syst} ~ \mbox{MeV} \nonumber \\
                                      & = & 4.21 ~ (26) ~ \mbox{MeV} ~ , 
\end{eqnarray}
which agrees with the experimental determination~\cite{ref:PDG}
\begin{equation}
    \left[ M_{\pi^+} - M_{\pi^0} \right]^{exp} = 4.5936 ~ (5) ~ \mbox{MeV}
    \label{e.pion_exp}
\end{equation}
within $\approx 1.5$ standard deviations.

\noindent In equation~(\ref{e.pion_splitting}), $()_{stat+fit}$ indicates the statistical uncertainty including also the one induced by the fitting procedure, while $()_{syst}$ is the total systematic uncertainty due to discretization effects, chiral extrapolation and finite volume effects.

Unlike the pion mass splitting, the mass difference $\left( M_{K^+} - M_{K^0} \right)$ is determined at the leading order by both e.m. and strong IB contributions. Adopting the quenched QED approximation and using the RM123 method, the result for the e.m. correction to the kaon mass splitting in the $\overline{MS}$ scheme at a renormalization scale equal to $\mu = 2$ GeV is
\begin{eqnarray}
  \label{e.kaon_qed_result}
    \left[ M_{K^+} - M_{K^0} \right]^{QED} (\overline{MS}, 2~\mbox{GeV}) & = & 2.07 ~~ (10)_{stat+fit} ~ (11)_{syst} ~\mbox{MeV} \nonumber \\
                                                                         & = & 2.07 ~~ (15) ~\mbox{MeV} ~ ,
\end{eqnarray}
where the systematic error also includes an estimate of the effects due to the quenched QED approximation.

Using the experimental value for the charged/neutral kaon mass splitting, $\left[ M_{K^+} - \right.$ $\left. M_{K^0} \right]^{exp} = -3.934 ~(20)$ MeV~\cite{ref:PDG}, one gets
 \begin{equation}
     \left[ M_{K^+} - M_{K^0} \right]^{QCD} (\overline{MS}, 2~\mbox{GeV}) = -6.00 ~ (15) ~\mbox{MeV}~ .
     \label{e.kaon_qcd}
 \end{equation}
By computing $\left[ \delta M_{PS} \right]^{IB}$ (see equation~(\ref{e.MPS_QCD})), it is then possible to evaluate the light-quark mass difference $\left( m_d - m_u \right)$ using equation~(\ref{e.kaon_qcd}). In reference~\cite{ref:Giusti:2017dmp} we found
\begin{equation}
    \left[ m_d - m_u \right] (\overline{MS}, 2~\mbox{GeV}) = 2.38 ~ (18) ~ \mbox{MeV} ~ ,
   \label{e.deltamud}
\end{equation}
which is consistent with the previous ETMC determination $2.67 ~(35)$ MeV~\cite{ref:Carrasco:2014cwa} at $N_f = 2+1+1$ and with the recent BMW result, converted in the ($\overline{MS}, 2~\mbox{GeV}$) scheme, $2.40 ~(12)$ MeV~\cite{ref:Fodor:2016bgu} at $N_f = 2+1$.

Finally, in reference~\cite{ref:Giusti:2017dmp} the evaluation of the $D$-meson mass splitting $\left( M_{D^+} - M_{D^0} \right)$ has been also addressed. Within the quenched QED approximation, the QED and QCD contributions to this are
\begin{eqnarray}
       \left[ M_{D^+} - M_{D^0} \right]^{QED} (\overline{MS}, 2~\mbox{GeV}) & = & 2.42 ~ (22)_{stat+fit} ~ (46)_{syst} ~ \mbox{MeV} ~ , \label{e.D_QED}
       \\[2mm]
       \left[ M_{D^+} - M_{D^0} \right]^{QCD} (\overline{MS}, 2~\mbox{GeV}) & = & 3.06 ~ (27)_{stat+fit} ~ (7)_{syst} ~ \mbox{MeV} ~ , \label{e.D_QCD}
\end{eqnarray}

Thus, combining the results (\ref{e.D_QED}) and (\ref{e.D_QCD}) the following prediction has been obtained
\begin{equation}
       M_{D^+} - M_{D^0} = 5.47 ~ (53) ~ \mbox{MeV} ~ ,
   \label{e.deltaMD}
\end{equation}
which is consistent with the experimental value $M_{D^+} - M_{D^0} = 4.75 ~(8)$ MeV \cite{ref:PDG} within $\simeq 1.4$ standard deviations.

\acknowledgments
I warmly thank my colleagues of the RM123 Collaboration for the enjoyable and fruitful work on the topics covered in this talk. I thank V. Lubicz for his comments on this manuscript. Work partially supported by the MIUR (Italy) under the contracts PRIN 2010 and PRIN 2015.

\end{document}